\documentclass{moriond}

\def\be{\begin{equation}}
\def\ee{\end{equation}}
\def\bea{\begin{eqnarray}}
\def\eea{\end{eqnarray}}

\newcommand{\Photo}{\includegraphics[height=35mm]{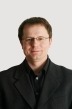}}

\begin{document}
\vspace*{4cm}
\title{SEARCH FOR A DARK PHOTON WITH THE WASA DETECTOR AT COSY}

\author{P. MOSKAL FOR THE WASA-AT-COSY COLLABORATION}

\address{Department of Physics, Jagiellonian University, Reymonta 4, Cracow, Poland}

\maketitle\abstracts{
We present recent results on the search for the U boson 
based on the data collected by means of the WASA detector 
and the Cooler Synchrotron COSY}

\section{Introduction}
Many astrophysical observations indicate existence of dark matter. 
One of the best example is delivered by the Chandra X-Ray Observatory 
which established that only a part of the colliding galactics (1E 0657-56)
emits X rays whereas the presence of remaining galactic mass can only be inferred based 
on the gravitational lensing~\cite{CHANDRA,CHANDRA-review}.
If the dark matter is utterly different from
the "Standard Model" matter then from the known interactions
it will feel only gravitational force.

Other experiments indicate that the present astrophysical models
cannot explain  magnitude and energy distributions of electrons and 
positrons~\cite{PAMELA,ATIC,FERMI,HESS} and 
a signal from  511 keV gamma quanta coming from the center of our Galaxy~\cite{INTEGRAL}. 
The origin of these phenomena may be explained assuming that positrons 
are created in the annihilation of the dark matter particles into  $e^+e^-$ pairs, 
and that this process is mediated by the U boson 
which may mix with the virtual photon~\cite{Boehm}. 
The existence of such a hypothetical boson 
with the mass in the order of 1~GeV
would affect the value of the branching 
ratios for the decays such as e.g.: $\pi^0 \to e^+e^-$ or $\eta \to e^+e^-$.  

The $\eta \to e^+e^-$ decay branching ratio has not yet been determined experimentally, and
on the basis of the Standard Model it is expected~\cite{Dorokhov} to be of the order of about $10^{-9}$. 
Low probability of this decay makes it sensitive to the hypothetical new forces 
that may indicate physics beyond the Standard Model. 
This is a very interesting opportunity 
to search for such effects at relatively low energies of the order of 1~GeV. 
The $\eta \to e^+e^-$ decay is not forbidden in the Standard Model but it has 
to proceed through an intermediate state with two virtual photons. Therefore, it is suppressed with 
respect to the $\eta \to \gamma\gamma$ decay by a factor $\alpha^2$ and the mass 
ratio $(m_e/m_\eta)^2$. Theoretical lower limit on the branching fraction 
BR($\eta\to e^+e^-$) $> 1.78 \times 10^{-9}$ results from the experimental 
value of the partial decay width $\Gamma(\eta\to\gamma\gamma)$~\cite{E1}. 

The interest in decays into lepton-antilepton pair is also  due to the results of the KTeV 
group which determined the value of BR($\pi^0\to e^+e^-$) = ( 6.44 $\pm$ 0.25 $\pm$ 0.22 ) $10^{-8}$~\cite{E3KTeV}, 
which is by about 3.3~$\sigma$ larger from the theoretical value obtained  based on the Standard Model~\cite{E4}. 
This result sparked speculations about a possible signature of physics beyond the Standard Model~\cite{E5,E6}. 

\begin{figure}
\centerline{a)\includegraphics[width=0.29\textwidth,clip=]{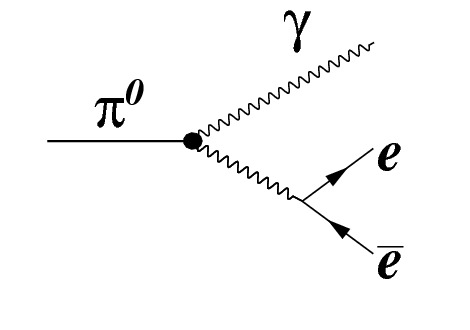}
b)\includegraphics[width=0.29\textwidth,clip=]{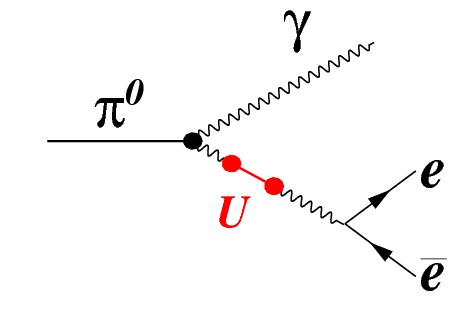}
c)\ \includegraphics[width=0.29\textwidth,angle=-90,origin=br,clip=]{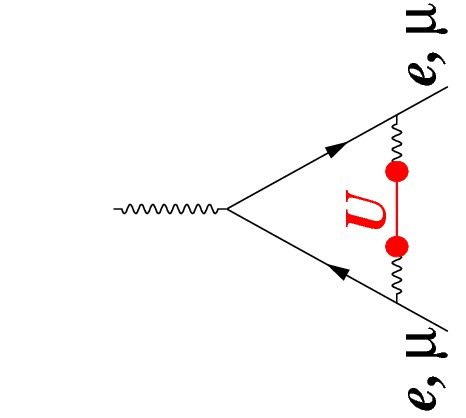}}
\caption[aa]{Feynman  diagrams for
a)  the lowest order electromagnetic $\pi^0\to e^+e^-\gamma$ decay
and possible  contributions of $U$ vector boson to: b)
$\pi^0\to e^+e^-\gamma$ and c) lepton $g-2$.
The figure and caption is adapted from reference~\cite{WASA_U}
\label{grafy}}
\end{figure}

The U boson would manifest itself also as 
a maximum in the $e^+e^-$ invariant mass distribution from the reactions such as  
e.g. $\eta \to e^+e^- \gamma$ or $\pi^0 \to e^+e^-\gamma$. 
Assuming the hypothetical coupling between the photon and the U boson  ($\gamma^*\to U$), 
such a decay can proceed via a following reaction chain:  
$\pi^0 \to  \gamma \gamma^* \to \gamma U \to \gamma e^+e^-$ 
as it is illustrated in Figure~\ref{grafy}.
Diagram presented in Figure~\ref{grafy}~c indicates a mechanism which may contribute to the g-2 anomaly
and therefore an existence of U boson may be also considerd as a possible source of 
discrepancy between
a g-2 value measured~\cite{Bennett:2006fi},
and predicted based on the Standard Model~\cite{SM-g2-1,SM-g2-2,SM-g2-3,SM-g2-4,SM-g2-5,KLOE-g2}.

WASA-at-COSY experiment has gathered the world largest  statistics of the $\eta$ and $\pi^0$ mesons. 
In this presentation a newest preliminary results obtained based on about 10\% of the data sample
are presented and discussed.

\section{The WASA-at-COSY experiment}

The design of the WASA detector has been optimized for the study of the 
$\eta\to e^+e^-$ and $\pi^0\to e^+e^-$ decays~\cite{WASA_HHA}. 
Therefore, data collected with this detector, 
together with the world's biggest statistics, 
create advantageous capabilities for studies of these rare decays. 
In addition, $\eta$ and $\pi^0$ mesons were produced in hadronic collisions 
near the threshold for their production which significantly reduces background 
of the electromagnetic processes. 

\begin{figure}
\begin{minipage}{0.9\linewidth}
\centerline{\includegraphics[width=0.75\linewidth]{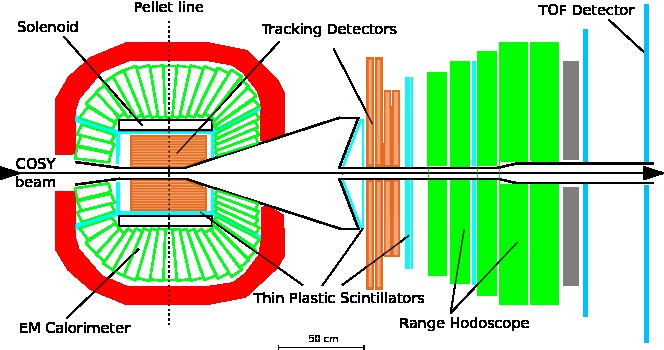}}
\end{minipage}
\caption[]{Scheme of the WASA detector setup installed at COSY accelerator}
\label{wasadetector}
\end{figure}

The WASA detector system (shown schematically in Figure~\ref{wasadetector}) consists of
the Forward Detector used for tagging of meson production, 
the Central Detector used for the registration
of the decay products, and the pellet target system.
The  proton-proton reactions occurs in the middle 
of the Central Detector in the intersection of COSY beam 
with the vertical beam of pellets. The interaction region is
surrounded by the multi-layer cylindrical drift chamber 
immersed in the axial magnetic field produced by the superconducting solenoid.
The outermost sensitive part of the Central Detector 
is the electromagnetic calorimeter covering 96 percent of the whole solid angle.
Particles flying in forward direction were registered 
in 14 scintillating layers and 16 layers of straw tubes detectors, 
while decay products of short-lived mesons $\eta$ and $\pi^0$ were measured in the cylindrical straw tube chamber, 
thin scintillator strips, and scintillating crystals of electromagnetic calorimeter. 
In order to identify the investigated 
reactions the technique of reconstruction of four-momenta of all particles in the final state 
and application of kinematic constraints are used. 
Leptons $e^+e^-$ and charged $\pi^+\pi^-$ mesons are 
identified based on the spectra of energy losses in the scintillator strips as a function 
of momentum determined from the trajectory curvature measured by means of straw tube chambers. 
Relation between the particle momentum and the energy deposited in the calorimeter is also applied. 
Gamma quanta produced in the $\pi^0$ meson decay are registered in the electromagnetic calorimeter 
and the identification of $\pi^0$ relays on the reconstruction of their invariant mass. 
Identification of protons and $^3$He ions scattered in forward direction is based on energy losses in the scintillating 
layers of forward detector and measurement of their trajectories in the straw tube chambers. 
After selection of events with appropriate particles in the final state the kinematic 
constrains are optimally used by application of kinematic fit with the methods 
of least squares and Lagrange multipliers. The $\chi^2$ test allows for optimal 
selection of investigated process and improvement of accuracy of four-momentum determination. 
Number of events corresponding to the production of desirable final state in the decay 
of $\eta$ or $\pi^0$ mesons is determined from the integration of the signal 
at the spectrum of the invariant mass of this set of particles.
As a result of experiments conducted with WASA-at-COSY we have collected a data sample of about 10$^9$ events 
with $\eta$ meson and about 10$^{11}$ with $\pi^0$ meson. 
These mesons were produced in the $pp \to pp\eta$ and $pp \to pp\pi^0$ reactions
where proton beam collided with hydrogen pellet target.

\section{Results}
Using the data collected with the WASA-at-COSY experiment for the $pp\to pp\pi^0$ reaction 
we have determined 
an  upper limit  
for the  square of  the
$U-\gamma$  mixing  strength   parameter  
$\epsilon^2$.  
Based on the 10\% of the collected statistics we set an upper limit of 
$5\times 10^{-6}$ at 90\% CL in the $M_U$ mass range from 20 MeV to 100~MeV.
This result 
significantly reduces the $M_U$   vs.  $\epsilon^2$  parameter
space which could explain the deviation between the
Standard Model prediction and the direct measurement of the anomalous
magnetic moment of the muon. For detailes the interested reader is referred to the 
recent WASA-at-COSY article~\cite{WASA_U}.

It is important to stress that 
currently, several experimental groups~\cite{KLOE_U1,KLOE_U2,APEX,BABAR,MAMI} 
carry out investigations searching 
for the signal from the U boson. This year  
a new more stringent result was published by the HADES collaboration~\cite{HADES_U}
and there are three other analysis reported on the arXive by the KLOE-2~\cite{KLOEnew}, MAMI~\cite{MAMInew} 
and BABAR~\cite{BABARnew} experiments.
These new results significantly reduce upper limits of the $\epsilon^2$ in the range between
20~MeV and 10~GeV leaving only a small space in the low mass range from 15~MeV to 30~MeV
in which a U boson can explain the discrepancy between predictions
based on the Standard Model and measurements of the g-2 muon anomaly. 
For other mass region this explanation is excluded.

Concerning the search for the $\eta \to e^+e^-$ signal 
the best published experimental upper limit was set by the HADES experiment 
BR($\eta \to e^+e^-$) $< 5.6 \times 10^{-6}$ at 90\% CL~\cite{E2HADES}.
So far the 
WASA-at-COSY collaboration, based on 5\% of the collected statistics,
has reported
a preliminary upper limit of $4.6 \times 10^{-6}$ at 90\% CL~\cite{berlowski}.  
The analysis of the remaining data sample is in progress. 

\section*{Acknowledgments}
We acknowledge support
by the Polish National Science Center through grants No. 0320/B/H03/2011/40, 2011/01/B/ST2/00431, 2011/03/B/ST2/01847,
by the FFE grants of the Research Center J\"{u}lich, by the Foundation for Polish Science (MPD programme),
by the EU Integrated Infrastructure Initiative HadronPhysics Project under contract number RII3-CT-2004-506078,
and by the European Commission under the 7th Framework Programme through the
’Research Infrastructures’ action of the ’Capacities’ Programme, Call: FP7~-~INFRASTRUCTURES~-~2008~-~1, 
Grant Agreement N. 227431.

\end{document}